\documentclass[twocolumn,prl,aps]{revtex4}

\def\duzomniejsze{<\kern-.7mm<}
\def\duzowieksze{>\kern-.7mm>}

\def\textbf#1{{\bf #1}}
\def\beq{\begin{equation}}
\def\eeq{\end{equation}}
\def\be{\begin{equation}}
\def\ee{\end{equation}}
\def\ben{\begin{eqnarray}}
\def\een{\end{eqnarray}}
\def\beqa{\begin{eqnarray}}
\def\eeqa{\end{eqnarray}}
\def\eea{\end{array}}
\def\bea{\begin{array}}
\newcommand{\bei}{\begin{itemize}}
\newcommand{\eei}{\end{itemize}}
\newcommand{\bee}{\begin{enumerate}}
\newcommand{\eee}{\end{enumerate}}

\def\hcal{{\cal H}}
\def\dcal{{\cal D}}

\def\bcal{{\cal B}}

\def\tr{{\rm Tr}}

\def\>{\rangle}
\def\<{\langle}
\def\blacksquare{\vrule height 4pt width 3pt depth2pt}

\def\ot{\otimes}

\def\wobie{\rightleftharpoons}

\def\tosym{S_{sym}}
\def\sym{{\cal S}_E^{as}}
\def\asym{{\cal A}_E^{as}}

\def\asymfin{{\cal A}_E}

\def\abba{S_{swap}}

\def\rateprzezsym{symmetry}
\def\rateswapa{swap-symmetry}

\newtheorem{lemma}{Lemma}

\newtheorem{proposition}{Proposition}
\newtheorem{theorem}{Theorem}
\newtheorem{definition}{Definition}

\def\bed{\begin{definition}}
\def\eed{\end{definition}}
\def\bel{\begin{lemma}}
\def\eel{\end{lemma}}
\def\bet{\begin{theorem}}
\def\eet{\end{theorem}}

\begin{document}

\title{Are quantum correlations symmetric ?}

\begin{abstract}
To our knowledge, all known bipartite entanglement measures  are
symmetric under exchange of subsystems. We ask if an entanglement measure that 
is not symmetric can exist. A related question is if there 
is a state that cannot be swapped by means of LOCC. 
We show, that in general one cannot swap states by LOCC.
This allows to construct nonsymmetric measure of entanglement, 
and a parameter that reports asymmetry of entanglement 
contents of quantum state. We propose asymptotic measure
of asymmetry of entanglement, and show that states for which 
it is nonzero, contain necessarily bound entanglement.
\end{abstract}

\author{ Karol Horodecki$^{(1)}$, Micha\l{} Horodecki$^{(2)}$, 
Pawe\l{} Horodecki$^{(3)}$}

\affiliation{$^{(1)}$Faculty of Mathematics Physics and Computer Science, 
University of Gda\'nsk, 80--952 Gda\'nsk, Poland}
\affiliation{$^{(2)}$Institute of Theoretical Physics and Astrophysics, University of Gda\'nsk, 80--952 Gda\'nsk, Poland}
\affiliation{$^{(3)}$Faculty of Applied Physics and Mathematics, Technical University of Gda\'nsk, 80--952 Gda\'nsk, Poland}

\maketitle

To our knowledge all known bipartite entanglement measures (EM) are symmetric 
under exchange of subsystems. To see it, it is enough to
to check whether objects entering  definitions of  a given measure 
are symmetric. For example, operational EMs like distillable entanglement $E_D$ \cite{BBPSSW1996}, 
and distillable key $K_D$ \cite{pptkey} are symmetric, because the sets
of target objects (maximally entangled states, private states) are symmetric, and 
the tools for distillation are the same for Alice and for Bob. 
The so called distance  EM's \cite{PlenioVedral1998} 
are symmetric, because the set of separable states is symmetric, and distance 
is invariant under unitaries, (which is even more general than symmetry under 
exchange of subsystems). 
As for the convex roof measures \cite{Vidal-mon2000,Uhlmann-roof}, they are 
defined by the measures of entanglement
on pure states. However {\it any} EM on pure state must be symmetric,
as it is function of eigenvalues of subsystem, which by Schmidt decomposition
are the same for both subsystems.

There are quantities related to entanglement, such as one-way distillable entanglement,
that are manifestly nonsymmetric \cite{BDSW1996}. However they are not true EM's in the sense 
that they can be increased by local operations and classical communication.  

Therefore we would like to rise the question: {\it Can entanglement be asymmetric?} 
In other words, can a state $\rho_{AB}$  have  more entanglement 
of some type than  the state $\sigma_{AB}= V\rho_{AB} V$  where $V$ is unitary operation 
that swaps subsystems? 

A closely related question is: {\it Can we swap a given state by LOCC?} Indeed,
if there exists a  measure that is nonsymmetric, it would increase under swap on some state, 
hence one couldn't swap it by LOCC. On the other hand, one can see that existence on 
states that are not swapable by LOCC leads to a nonsymmertic EM. At first, the question
may seem trivial: it is known that swap is highly nonlocal gate, which 
if applied to halves of singlets (produced locally), can  create 2 bits of 
entanglement. However, the other halves of singlets stay untouched:
we act with swap on two halves and with {\it identity} on two other halves. 
Swap {\it itself} cannot create entanglement out of separable states
(cf. \cite{Busch-swap,ZanardiZF-ep2000}), because $V\psi\ot\phi=\phi\ot \psi$. 

In this paper we will show that indeed, entanglement can feel where is left and 
where right-hand-side of the system. More specifically, we will first show that 
in general, one cannot swap states by LOCC. We will then exhibit an 
asymmetric measure of entanglement. Interestingly, the impossibility of swapping 
we will proven by use of a usual symmetric EM. 
 
We then consider asymptotic setup, and conjecture, that one cannot swap 
even in the regime of many copies, and allowing for (asymptotically vanishing) error. 
We define two asymptotic symmetry/asymmetry measures  and show that they 
coincided. Moreover we exhibit connection between asymmetry and bound entanglement:
if there is nonzero asymmetry of entanglement of a given state, 
then the state necessarily contains bound entanglement. We give also quantitative 
relations between asymmetry and bound entanglement contents.

{\it Existence of non-swapable states.}  

We will now prove that for a large class of states 
one can swap them by LOCC 
only when one can swap them by local unitaries.  The class  consists 
of all states that have full Schmidt rank  \cite{Terhal-Pawel-rank}.
Equivalently, such states can be characterized by a measure of entanglement 
introduced by Gour \cite{Gour-mon2004}. The measure 
for pure state is given by 
\be
G(\psi_{AB} ) =d (\det \rho_A )^{1\over d}
\ee 
where $\rho_A$ is reduced density matrix of $\psi$. For mixed states $G$ is given by
\be
G(\rho)=\inf \sum_i p_i G(\psi_i)
\ee
where infimum is taken over decompositions $\rho=\sum_i |\psi_i\>\<\psi_i|$.
(This is standard convex roof procedure \cite{Vidal-mon2000,Uhlmann-roof}.)
Note that $G(\psi)$ is nonzero if and only if $\psi$ has maximal Schmidt rank. 
It follows that our class  of mixed states is characterized by $G(\rho)>0$.
Thus, we will prove that  if $G>0$, then swapping by LOCC means swapping by product unitary.

In particular, it follows that if state with $G>0$ has different entropies 
of subsystems, it cannot be swapped by LOCC, since clearly local unitaries cannot 
change local entropy. Moreover, for two-qubit states, 
the condition $G>0$ is equivalent to entanglement so that we obtain 
that any two qubit state is LOCC swapable iff it is swapable by $U_A\ot U_B$.

Our main result is contained in the following theorem
\bet
\label{thm:G}
Consider state $\rho$ acting on $C^d\ot C^d$, for which $G>0$ (equivalently, with 
Schmidt rank equal to $d$). Then, if such state can be swapped by LOCC, 
then it can be also swapped by some product unitary operation $U_A \ot U_B$.
\eet

To prove this theorem we need two lemmas. 

\bel
\label{lem:convex}
For any state $\rho$ on  $C^d \ot C^d$, and trace preserving separable operation 
$\Lambda(\cdot)=\sum_i A_i \ot B_i (\cdot) A_i^\dagger \ot B_i^\dagger$ there holds
\be
\sum_ip_i G(\sigma_i) \leq \sum_i |\det A_i|^{1\over d}  |\det B_i|^{1\over d} 
G(\rho)
\ee
where $\sigma_i= {1\over p_i} A_i \ot B_i (\rho) A_i^\dagger \ot B_i^\dagger$,
$p_i=\tr (A_i \ot B_i (\rho) A_i^\dagger \ot B_i^\dagger)$.
\eel
{\it Remark.} Similar result (with equality) was obtained for concurrence in 
\cite{VerstraeteDM-det2001}. In the proof we will use, in particular, techniques  from the proof of 
monotonicity of convex roof EM's under LOCC \cite{Vidal-mon2000,Michal2001}.

{\bf Proof.} 
Consider optimal decomposition $\rho=\sum_j q_j |\psi_j\>\<\psi_j|$,
so that $G(\rho)=\sum_jq_j G(\psi_j)$. One finds that 
\ben
\sigma_i=\sum_j {q_j p_i^{(j)}\over p_i} \left({1\over p_i^{(j)}} {X_i|\psi_j\>\<\psi_j|X_i^\dagger}\right)\\
\equiv \sum_j r_j^{(i)} |\phi_j^{(i)}\>\<\phi_j^{(i)}|
\een
where we have denoted $X_i=A_i\ot B_i$, $p_i^{(j)}= 
\tr (X_i |\psi_j\>\<\psi_j| X_i^\dagger)$. The coefficients $r_j^{(i)}$ are probabilities for
fixed i 
 and $\phi_j^{(i)}$ are normalized states. 
We then have 
\ben
&&\sum_i p_i G(\sigma_i)= \sum_i p_i G(\sum_j r_j^{(i)} |\phi_j^{(i)}\>\<\phi_j^{(i)}|) \leq \nonumber\\
&&\leq \sum_{ij} p_i r_i^{(j)} G(\phi_j^{(i)})=\sum_{ij} q_j G(X_i \psi_j)
\een
where we have used convexity of $G$ and the fact that $G(\alpha\rho)=\alpha G(\rho)$ for 
$\alpha\geq 0$.
Now, as shown in \cite{Gour-mon2004} $G(A\ot B \psi)= |\det A|^{1\over d}|\det B|^{1\over d}$.
It follows that 
\ben
&&\sum_i p_i G(\sigma_i) \leq \sum_i|\det A_i|^{1\over d}|\det B_i|^{1\over d} =
\sum_j q_j G(\psi_j)\nonumber \\
&&\sum_i|\det A_i|^{1\over d}|\det B_i|^{1\over d}G(\rho)
\een
This ends the proof of the lemma. \blacksquare

The second lemma we need is as follows
\bel
\label{lem:aibi}
For operation $\Lambda$ from lemma \ref{lem:convex}
we have $\sum_i|\det A_i|^{1\over d}|\det B_i|^{1\over d}\leq 1 $
with equality if and only if $\Lambda$ is mixture of product unitary operations.
\eel
{\bf Proof.} Note that  $|\det A_i|^{1\over d}|\det B_i|^{1\over d} = [det (X_i^\dagger X_i)]^{1\over d^2}$ where $X_i=A_i \ot B_i$. We then have 
\be
[\det (X_i^\dagger X_i)]^{1\over d^2} \leq {1\over d} \tr (X_i^\dagger X_i)
\ee
as this is actually the inequality between geometric and arithmetic mean 
of eigenvalues of $X_i^\dagger X_i$ (cf. \cite{Gour-ass2005}). It then follows that equality can hold  if and only if 
all eigenvalues are equal i.e. when $X_i^\dagger X_i$ is proportional to identity.
Summing up we get 
\be
\sum_i |\det A_i|^{1\over d}|\det B_i|^{1\over d} \leq  
{1\over d^2} \tr \sum_i(X_i^\dagger X_i) = 1
\ee
where used the fact that $\Lambda$ is trace preserving, so that $\sum_i X_i^\dagger X_i=I$. 
Equality can hold
only when it holds for all terms, which implies that 
$(A_i\ot B_i)^\dagger (A_i \ot B_i)$ 
is proportional to identity. Hence  $A_i$ and $B_i$ are proportional to unitaries. 
Thus, $\Lambda$ is mixture of product unitary operations. \blacksquare

{\bf Proof of the theorem \ref{thm:G}}.
We assume that $G(\rho)>0$ and that we can swap $\rho$ by LOCC,
i.e. $\Lambda(\rho)=V \rho V$. We will now use notation from the lemmas.   
Thus we assume that $\sum_ip_i \sigma_i=V \rho V$. 
Using invariance of $G$ under swap, convexity of $G$ and lemma \ref{lem:convex}
we obtain
\ben
&&G(\rho)=G( V\rho V ) = G(\sum_i p_i \sigma_i)\leq \sum_ip_i G(\sigma_i) \leq \nonumber\\
&&\leq
\sum_i |\det A_i|^{1\over d}|\det B_i|^{1\over d} G(\rho) 
\een
Since $G(\rho)>0$ we get  $\sum_i |\det A_i|^{1\over d}|\det B_i|^{1\over d}\geq 1$.
Thus in view of lemma \ref{lem:aibi} we obtain that $\Lambda$ 
must be mixture of product unitaries:
\be
\Lambda(\rho)=\sum_ip_i U_A^i \ot U_B^i \rho U_A^{i\dagger} \ot U_B^{i\dagger}
\equiv\sum_i p_i \sigma_i
\ee
Then the states $\sigma_i$ have the same von Neumann entropy $S$ as $\rho$, so that 
\be
S(\sum_ip_i \sigma_i) = S( V \rho V) = S(\rho)= \sum_ip_i S(\sigma_i) 
\ee
Now, from strict concavity of entropy we obtain that all $\sigma_i$'s must be the same, 
so that $V \rho V = U_A^1 \ot U_B^1  \rho U_A^{1\dagger} \ot U_B^{1\dagger}$.
Thus swap can be performed by local unitary operation.  \blacksquare

{}{\it Examples.} From the theorem it follows that all entangled 
two qubit states are swapable, if they are swapable by $U_A\ot U_B$ 
Thus any state with subsystems of different spectra is not LOCC swapable,
since local unitaries keep local spectra. Exemplary state is mixture of $|01\>$ and 
${1\over \sqrt{2}} (|00\> + |11\>)$. 

Let us see, whether the assumption that $G>0$ is essential. For higher dimensions there 
are many states that have $G=0$. One would be tempted to think that 
for any entangled state that is LOCC swapable, we can swap it by local unitaries.
However, it is not true. Consider state on $C^2\ot C^4$ system:
being a mixture of $\psi_+={1\over \sqrt 2} (||00\>+|11\>)$ and 
$\psi={1\over \sqrt 2} (||02\>+|13\>)$.
The subsystems have different spectra, so that we cannot swap it by local unitaries. 
However, the mixture can be reversibly transformed into e.g. $\psi_+$ by local unitary. 
Thus it can  be swapped.

{}{\it Asymmetric EM.}
We take any "distance" $\dcal$  which is continuous, satisfies 
$\dcal(\Lambda(\rho)|\Lambda(\sigma))\leq \dcal(\rho,\sigma)$ and $\dcal(\rho,\sigma)=0$ 
if and only if $\rho=\sigma$. 
We consider associated measure $E^\dcal(\rho)=\inf_{\sigma_{sep}}\dcal(\rho,\sigma_{sep})$ 
\cite{PlenioVedral1998} where infimum is taken over all separable states. 
Consider then a fixed state $\sigma$ that cannot be swapped by LOCC. 
Now, our measure is defined as 
\be
E_{\sigma}(\rho )=E^\dcal(\sigma) - \inf_\Lambda\dcal(\sigma,\Lambda(\rho))
\ee
where infimum is taken over all LOCC operations $\Lambda$. 
Note that for separable states $E_{\sigma}=0$, and that by definition it does not 
increase under LOCC. We have $E_{\sigma}(\sigma)=E^\dcal(\sigma)$ while 
$E_{\sigma}(V\sigma V)<E^\dcal(\sigma)$. To see it note that 
if we cannot swap a state exactly, then we also cannot swap it with arbitrary 
good accuracy according to distance satisfying the above conditions. 
This follows from compactness of set of separable operations. Thus the 
second term is nonzero. 

{}{\it Measure of asymmetry of entanglement.}
We can define a parameter that would report asymmetry of entanglement 
of a given state. 
\be
\asymfin(\rho)= \inf_\Lambda \dcal(\Lambda(\rho), V \rho V )
\ee
where infimum is taken over all LOCC operations $\Lambda$. Clearly, it is nonzero 
if and only if a state cannot be swapped by LOCC.

{\it Asymptotics.}
So far we have talked about exact transformations. It is interesting to ask if the 
effect survives limit of many copies, where we allow inaccuracies that vanish 
asymptotically. We have not been able to answer this question, however we 
think it is most likely, that even asymptotically, in general one 
cannot swap states by LOCC.

Under such assumption, we can consider a
parameter, which will report {\it asymptotic symmetry} of entanglement.

To define this parameter we need the notion 
of optimal transition rate of given state $\rho$ to other state $\sigma$ 
denoted as $R(\rho\rightarrow \sigma)$ which is the maximal
ratio $m\over n$ of the transformation $\rho^{\ot n}\rightarrow \sigma' \approx 
\sigma^{\ot n}$ via some LOCC map \cite{Michal2001}.

{\definition Let $\rho_{AB} \in \bcal(\hcal_A\ot\hcal_B)$ be an entangled
state.  
Then \rateswapa\ is defined for entangled
states as follows:
\be
\abba(\rho) = R(\rho \rightarrow V\rho V^{\dagger}).
\ee
which is the optimal rate of transition from $\rho$ to $V\rho V$ by
means of LOCC.
}

This quantity is clearly infinite for separable states. 
However for entangled states it is always finite
\bel
For entangled state $\rho$ we have 
\be
\abba(\rho)\leq 1
\ee
\eel
{\bf Proof.}
We apply relation between rates and asymptotically continuous entanglement 
monotones \cite{Michal2001}. Consider two state $\sigma$ and $\rho$, and an asymptotically 
continuous  entanglement monotone $E$. Let us assume that $E^\infty(\sigma)> 0$. 
Then we have 
\be
R(\rho\to \sigma) \leq {E^\infty (\rho) \over E^\infty (\sigma)}
\ee
Here we will take $\sigma=V \rho V$ and $E$ to be entanglement of formation $E_F$. 
Regularization of $E_F$ is entanglement cost: $E_F^\infty=E_c$  and it was shown 
in \cite{YangHHS2005-cost} that it is nonzero for any entangled state.
Since $E_c(\rho)=E_c(V \rho V)$ we obtain that $R(\rho\to V\rho V )\leq 1$
which ends the proof. \blacksquare

We can also design another quantity, which would  also 
report how much asymmetric is entanglement
of a given state.  To this end let us consider round-trip-travel 
rate  i.e. the optimal rate of transferring state $\rho$ into itself via 
some other state $\sigma$ (cf. \cite{MichalSS2002}). It is formally
defined as 
\be
R(\rho\wobie\sigma)= R(\rho\to \sigma) R(\sigma\to \rho)
\ee
where we use convention $0 \cdot \infty=\infty \cdot 0=0$.
We now define our second quantity:
\bed
The following quantity 
\be 
\tosym(\rho)=\sup_{\sigma}R(\rho\wobie \sigma)
\ee 
where supremum is taken over all symmetric states $\sigma$
we will call {\rm \rateprzezsym}.
\eed
Again, using \cite{YangHHS2005-cost} we can get that 
for any entangled state $\tosym\leq 1$. 

However surprisingly, it turns out that the two quantities are equal:
\begin{proposition}
The quantities $\tosym$ and $\abba$ 
are equal to each other
\be
\tosym=\abba
\ee
\end{proposition}

{\bf Proof.} To see that $\tosym\leq \abba$ 
consider the protocol achieving $\tosym$
\be
\rho^{\ot n} \to  \sigma^{\ot m} \to \rho^{\ot k}
\ee
where $\sigma=V\sigma V$. Since the protocol is optimal, 
we have $k/n  \approx \tosym$. 
In the second stage (transforming $\sigma$ into $\rho$ 
let us exchange roles of Alice and Bob. Then, instead of 
$\rho^{\ot k}$ we will obtain $(V \rho V)^{\ot k}$.
Thus the total protocol will simply swap the state with 
rate $k/n$. Thus we can swap at least with rate $\tosym$ 
which proves $\abba \geq \tosym$.  

To prove converse,  it is enough to find a symmetric state $\sigma$ 
such that $R(\rho\wobie \sigma)$ will be equal to $\abba$. 
Clearly, instead of symmetric (i.e. swap invariant state) we can 
choose a state which can be made symmetric by local unitaries. 
We will take 
\be
\sigma= \rho \ot  V\rho V
\ee
It is easy to see that local swaps produce a symmetric state from $\sigma$.
We will now express $R(\rho \wobie \rho\ot V\rho V)$ 
in terms of $\abba(\rho)$. 
To this end consider the following transformation 
\be
\rho^{\ot n}\ot \rho^{\ot m}  \to (V\rho V)^{\ot m} \ot \rho^{\ot m} =\sigma^{\ot m}
\ee
where the rate $m/n\approx\abba$ is possible by definition of $\abba$.  
Then we consider transformation  that returns to the state $\rho$:
\be
\sigma^{\ot m}= (V\rho V)^{\ot m} \ot \rho^{\ot m} \to \rho^{\ot k} \ot \rho^{\ot m}
\ee
where again by definition of $\abba$ the rate $k/m\approx \abba$ is possible. 
Thus the overall round-trip-travel rate vis state $\sigma$ satisfies
\be
R(\rho\wobie \sigma)\leq {k+m\over n+m} \approx  {\abba +1 \over {1\over \abba}  + 1} 
= \abba 
\ee
Since $\tosym$ is supremum of such rates, we obtain that $\tosym \geq \abba$. 
This ends the proof.
\blacksquare

We thus obtain our asymptotic quantities measuring symmetry/asymmetry.
\bed
The quantity  $\tosym=\abba$ we will call symmetry of entanglement, and 
will denote by $\sym$. The quantity $\asym=1-\sym$ we will call asymmetry of entanglement.
\eed

Thus, entanglement in a given state is not symmetric 
when $\asym >0$. We will now argue that  states with nonsymmetric entanglement 
must possess bound entanglement, i.e. for such state distillable entanglement 
is strictly smaller than entanglement cost $E_D< E_c$. Thus asymptotic asymmetry 
brings irreversibility. The reason is obvious, reversibility in 
distillation-creation process means that we can go reversibly from $\rho$ 
to $\rho$ through maximally entangled state
which is symmetric state. Thus $\sym=1$ in such case. We have

{\theorem  For entangled states, we have 
\be
{E_D\over E_c} \leq \sym \leq 1
\ee
Equivalently we have 
\be
{E_b \over E_c} \geq \asym
\ee
where $E_b=E_c-E_D$.
}

{\bf Proof.} The optimal rate  $R(\rho\wobie \psi_+)$ 
where $\psi_+={1\over 2} (|00\>+|11\>)$ is given by 
\be
R(\rho\wobie \psi_+) = {E_D\over E_c}
\ee
Since maximally entangled state is symmetric,
this is rate of a particular protocol of round-trip-travel 
from $\rho$ to $\rho$ via symmetric state. Thus it is no greater than 
$\sym$ which is supremum of rates over such protocols. \blacksquare

From this theorem it follows that $\sym$ is nonzero for distillable states.

{\it  Concluding remarks.}

In this paper we propose a measure of asymmetry of entanglement
for a single copy of quantum state. This proposition is not unique. Other candidate
can be the infimum of distance from the set of single copy LOCC swapable
states. It appears that the lower bound on this measure in terms of $G$-concurrence
can be found.

We also conjecture that entanglement can be asymmetric in assymptotic regime
of many copies i.e. that there exist states with $\sym < 1$.
One could then ask if $V$ can increase $E_D$ of some 
distillable states i.e. if $E_D(\rho \ot V\rho V) > E_D(\rho^{\ot 2})$.

If however $\sym =1$ for all states one would have that certain nontrivial
task can be achieved via LOCC. Moreover the nice correspondence between
transposition and swap would hold. 
As we have mentioned, like $I\ot T$ is not physical, 
the operation $I\ot V$ can not be implemented by means of LOCC
i.e. it is not physical with respect to this class of operations.
Although transposition is not completely positive it can be 
performed on a {\it known} state, as it is positive.  If then
$\sym =1$ for all states i.e. all states would be swapable, then V
like T could be performed on a {\it known} state (in this case via LOCC operations).

Note that still there are many states which have $\sym=1$ because they 
are swap invariant. It is then tempting to develop a scheme of symmetry of 
entanglement with respect to certain group G of unitary transformations 
(see in this context \cite{Group-asym05} and \cite{WernerVollbrecht}).
That is $G-$symmetry of a state would be maximal rate of distillation of 
states which are invariant under actions of $G$. 

As a generalization of our approach one can consider the asymmetry 
of general quantum correlations by restricting class of allowed operations
to so called {\it closed} LOCC operations \cite{huge-delta}. 
In such case also certain separable states may exhibit asymmetry. 
Moreover in analogy to asymmetry of entanglement one 
can also quantify asymmetry of private (cryptographic) correlations.

Finally, we note that quite recently
other interesting investigations of notion of exchange of subsystems 
and swap symmetry have been independently developed 
\cite{JonoW-uncommon2005,Harrow-Shor2005,Linden-Smolin-Winter05}.

\begin{acknowledgments}
We would like to thank Ryszard Horodecki, Aditi Sen(De) and Ujjwal Sen
for helpful discussion. Part of this work has been done while the authors 
took part in the QIS Programme in Isaac Newton Institute (Cambridge),
which hospitality is gratefully acknowledged.
This work is supported by MNiI, grant PBZ-MIN-008/P03/2003,
EU grants RESQ (IST-2001-37559), 
QUPRODIS (IST-2001-38877) and 
EC IP SCALA.  
\end{acknowledgments}

\bibliographystyle{apsrev}
\bibliography{C:/1odzyskane/dysk_d/PRACE/Referencje/refpostmich,C:/1odzyskane/dysk_d/PRACE/Referencje/refmich}

\end{document}